\documentclass[aps,twocolumn,showpacs,nofootinbib]{revtex4-1}
\usepackage{graphicx}   %       for graphics
\usepackage{latexsym}   %       for special symbols
\usepackage{enumerate}
\usepackage{color}
\usepackage{ulem}
\usepackage{float}

\begin{document}

\title[]{Susceptibilities of strongly interacting matter in a finite volume}

\author{Christian Spieles$^{1}$, Marcus Bleicher$^{1,2,3,4}$ and Carsten Greiner$^2$}

\affiliation{$^1$~Frankfurt Institute for Advanced Studies, Ruth-Moufang-Strasse 1, D-60438 Frankfurt am Main, Germany\\
             $^2$~Institut f\"ur Theoretische Physik, Goethe-Universit\"at, Max-von-Laue-Strasse 1, D-60438 Frankfurt am Main, Germany\\
             $^3$~GSI Helmholtzzentrum f\"ur Schwerionenforschung, Planckstr. 1, D-64291 Darmstadt, Germany\\
             $^4$~John von Neumann-Institut f\"ur Computing, Forschungszentrum J\"ulich, D-52425 J\"ulich}

\email{spieles@fias.uni-frankfurt.de\\ bleicher@th.physik.uni-frankfurt.de\\ carsten.greiner@th.physik.uni-frankfurt.de}

%\date{\today \quad - DRAFT}

\begin{abstract}
We investigate possible finite-volume effects on baryon number
susceptibilities of strongly interacting matter. Assuming that
a hadronic and a deconfined phase both contribute to the thermodynamic state
of a finite system due to fluctuations, it is found that the resulting
shapes of the net-baryon number distributions deviate significantly from the
infinite volume limit for a given temperature $T$ and baryochemical potential
$\mu_B$. In particular, the constraint on color-singletness
for the finite quark-gluon phase contribution leads to a change of the
temperature dependence of the susceptibilities in finite volumes.
\textcolor{black}{According to the model, the finite-volume effect depends
qualitatively on the value of $\mu_B$.}
\end{abstract}

\maketitle

\section[]{Introduction}
In the last years considerable attention has been devoted to event-by-event fluctuations and correlations of conserved quantities in 
relativistic heavy-ion collisions (see \cite{Luo2017} for an overview of the
beam energy scan program at RHIC and the theoretical background).  
One important goal of this endeavor is to experimentally probe the phase structure
of strongly interacting matter and possibly identify its critical point.
Direct comparisons of measured net-baryon cumulants with the
corresponding thermodynamic susceptibilities from lattice QCD calculations  \cite{Bazavov2017}\cite{Guenther2018} appear
very promising in this respect. These observables are believed to provide rather robust
signatures of the underlying thermodynamics in the case of heavy-ion
experiments.
However, in order to relate experimental measurements and
theoretical equilibrium properties, several complications have to be
considered:
One practical problem is the finite phase space acceptance and the limited efficiency of any
real experiment which may obscure relevant signatures in the data. Necessary corrections have been addressed
theoretically, e.~g. in \cite{Bleicher2000}\cite{Kitazawa2012}\cite{Alba2014}\cite{Bzdak2015}. 
A fundamental and obvious issue is the fact that conserved charges do not
fluctuate globally. Therefore, one has to restrict the statistical analysis to a
part of the total phase space of the heavy-ion reaction. The acceptance window
of such an analysis must be chosen
sufficiently small, so that the observables can be considered as reflecting a subsystem in thermodynamic
contact with a heat bath. Only then, the grand canonical
ensemble can be applied to the theoretical description of this subsystem. The potential influence of a finite volume of the source has been explored in  
\cite{Nahrgang2012}\cite{Redlich2016}\cite{Karsch2016}\cite{Jeon2003}. At the same time,
however, the subsystem must be large enough to allow the relevant particle correlations to show
in the first place, see e.~g. \cite{Koch2016}. Finally, one has to take into
account initial state (e.~g. volume) fluctuations in
heavy-ion experiments, which overlay the possibly critical fluctuations
under investigation \cite{Skokov2013}. 

In the following, we want to draw the
attention to an aspect, which has not been addressed
yet: Even under ideal circumstances and with complete control over
all the beforementioned issues, a heavy-ion experiment can only probe finite
volumes of strongly interacting matter, while lattice QCD calculations
generally refer to the infinite volume case. In a finite volume, the partition
function of the system is subject to additional constraints which may lead
to changes in the thermodynamic properties like the baryon number susceptibilities.
Firstly, in a finite volume we expect fluctuations of the phase composition that are suppressed in the infinite volume case:
I.~e. admixtures of unfavourable macroscopic configurations of
the system can be realized with finite probability and can have an effect 
on the thermodynamics of the system. Secondly, when the finite hadronic system of a heavy-ion
reaction is forced to undergo a phase transition to a deconfined phase by the collision
dynamics, we expect the finite quark-gluon plasma phase to be suppressed -  as
compared to the case of infinite matter - due to the
requirement of color-singletness, i.~e. an explicit volume dependence of the quark-gluon plasma equation of
state. In this letter, we present an exploratory study of these effects and their implications with respect to net-baryon fluctuations 
for scenarios with different baryo\-chemi\-cal potentials.

\section[]{The model}\label{sec:model}

We extend the model proposed in \cite{Spieles1998} to allow for the investigation of
finite-volume effects on baryon number
susceptibilities as function of temperature and chemical potential.
The basic assumption of the schematic model is a first order phase transition
between a hadronic phase and a quark-gluon plasma phase, which is justified for sufficiently high values of the baryochemical potential.
\footnote{For $\mu_B=0$, it is not in accordance with lattice QCD
calculations. However, we expect that the qualitative behaviour of the addressed effects will also be present - although less pronounced - in the case of a crossover.}
In a finite volume, it presumes coexistence of the two phases due to fluctuations: As a consequence of the finite
(fixed) volume of the total system, any macroscopic configuration $\bf x$
contributes with a probability  $p({\bf x}) \sim \exp [-\Phi({\bf x})/T]$,
where $\Phi(\bf x)$ is the grand canonical potential of the system
\cite{Landau1976}. In the present simplified set-up, the macroscopic configuration is the composition
of the total system in terms of the partial volumes of the individual phases,
which are assumed to be microscopically uncorrelated, i.~e., the partition function
factorizes. In this picture, all intensive thermodynamic quantities of the total
macroscopic system are given as expectation values based on the weight of  all possible
configurations.

The equations of state of the subsystems are a relativistic ideal quantum
gas of non-strange hadron resonances (with eigenvolume correction) on the one
hand, and a relativistic ideal quantum gas of two massless quark flavors and gluons
confined in an MIT bag on the other hand.

For further details we refer the reader to the appendix.

\section[]{Susceptibilities of baryon number in the finite system}

The susceptibilities of the baryon number in the finite system are calculated as 
\begin{equation}
\chi^B_i = -\frac{\partial ^i \hat{\varphi}}{\partial \hat{\mu}^i_B} \quad ,
\end{equation}
from the dimensionless density of the grand canonical potential $\hat{\varphi}=
\Phi (T,\mu_B,V) V^{-1} T^{-4}$, where $\hat{\mu}_B =\mu_B/T$ is the reduced baryochemical potential.

The first order susceptibility is proportional to the expectation value of
the net-baryon number, 
\begin{equation}
\chi^B_1(V) V T^3= <N_B> \quad .
\end{equation}

For $\mu_B=0$ this quantity has a value of zero, which, of course, is reflected
by the model outcome.
Let us consider the second order susceptibility, which
is proportional to the variance of the net-baryon number:

\begin{equation}
\chi^B_2(V) V T^3 = \sigma_B = <(\delta N_B)^2> \quad .
\end{equation}

\begin{figure}[h]
\includegraphics[width=0.5\textwidth]{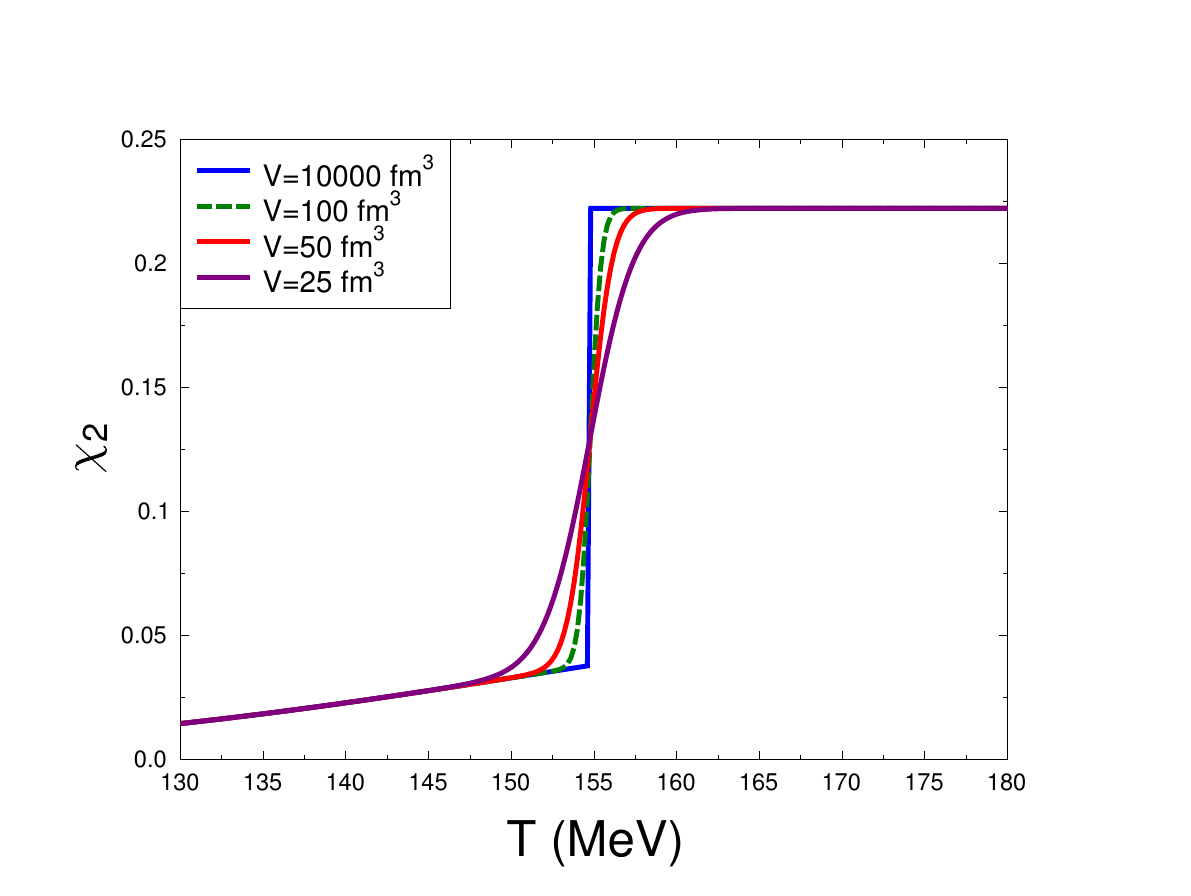}
\caption{
Second order baryon number susceptibilities $\chi^B_2$ as function of
temperature at $\mu_B=0$. For the individual phases  the
equations of state without explicit volume effects are employed.
} 
\label{fig:chi2inf} 
\end{figure}

Figure \ref{fig:chi2inf} shows this quantity as a function of temperature at
$\mu_B=0$ for different system volumes. The quark-gluon plasma phase is
constructed without color-singlet and zero-momentum constraint, i.~e., there
is no explicit volume dependence. The model calculations show 
a softening and broadening of the phase transition which is more pronounced for small volumes.
As was discussed in \cite{Spieles1998}, this can be understood as a
consequence of the admixture of the ``unfavourable'' phase in the finite system at any given
temperature. Although suppressed exponentially, the presence of the
quark-gluon phase below the critical temperature $T^{\infty}_C$ and the presence of
the hadronic phase above $T^{\infty}_C$ has a finite probability ($T^{\infty}_C$ marks
the first order phase transition for infinite matter). The effect
presented in Figure~\ref{fig:chi2inf} should in principle be relevant in any realistic
model-treatment of strongly interacting matter in finite volumes. The basic infinite matter
equations of state of the phases used in our model and the resulting
equation of state of the two-phase system are of course only schematic approximations to study the finite size effects, as can be seen from the comparison
with lattice QCD results\footnote{For $\mu_B=0$, lattice QCD does not exhibit a discontinuity of 
$\chi^B_2(T)$ for infinite matter as our simple model, but rather a smooth
transition.}. However, the model calculation gives an indication of how strong the deviation from a given baseline -
the true properties of infinite matter - can be due to the finite volume.
In this respect one would conclude from Fig.~\ref{fig:chi2inf} that the
finite volume effect considered here should in principle affect the observed slope of the crossover
curve of $\chi^B_2(T)$. But the effect seems too small to be relevant for the analyses 
of heavy-ion experiments at present. 

\begin{figure}[h]
\includegraphics[width=0.5\textwidth]{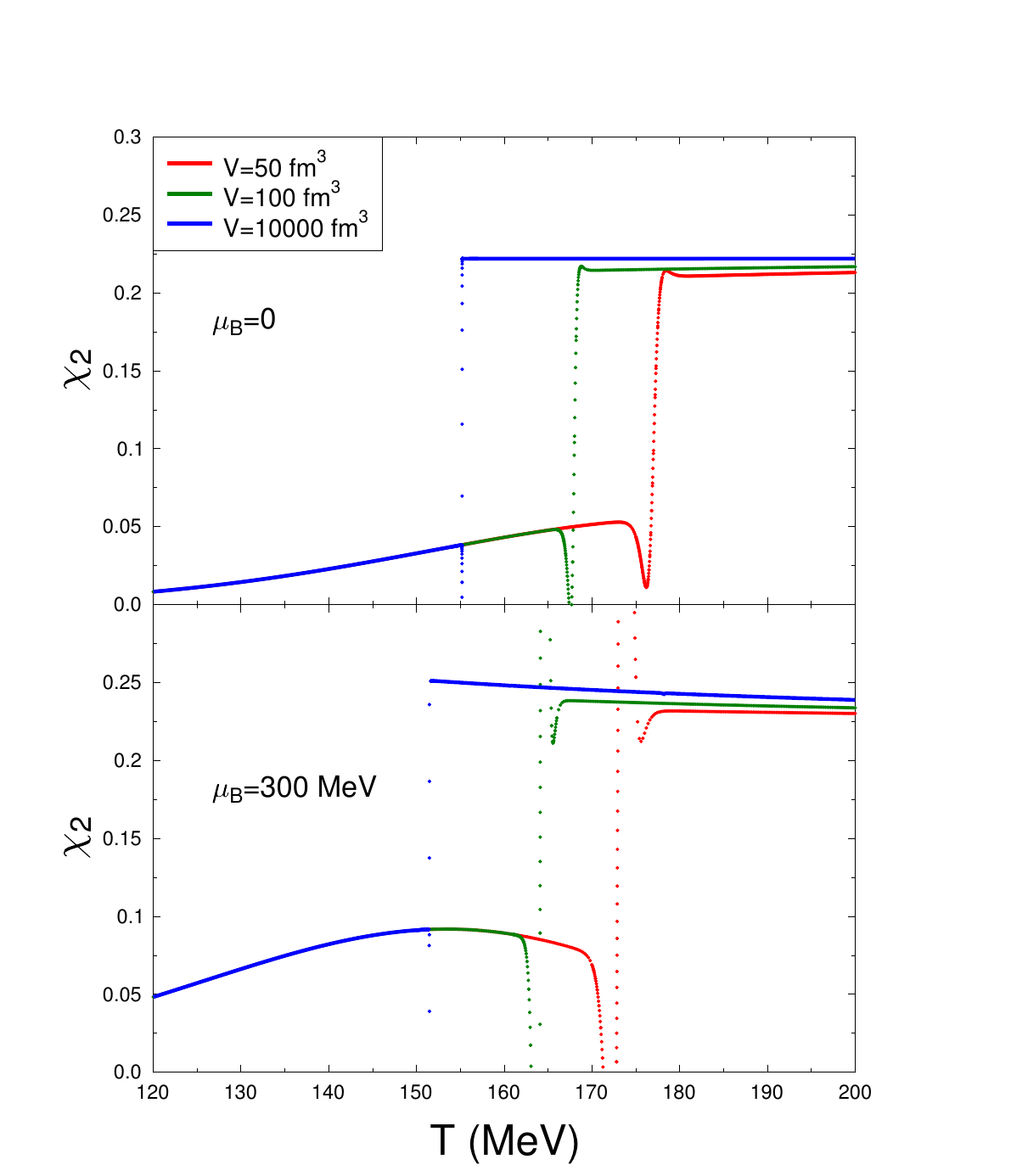}
\caption{
Second order baryon number susceptibilities $\chi^B_2$ as function of
temperature for different system volumes at $\mu_B=0$ (top) and $\mu_B=300~{\rm MeV}$ (bottom). For the QGP phase the
equation of state with explicit volume effects is employed.
} 
\label{fig:chi2} 
\end{figure}

Figure~\ref{fig:chi2} (top) shows the same quantity as Figure~\ref{fig:chi2inf}
for the two-phase model, now explicitely taking into account the color-singlet constraint
in the quark-gluon equation of state. The picture now changes completely. The
softening of the step in the second order susceptibility is reduced, while
the temperature of the sudden rise is shifted drastically with decreasing
volume. Both effects are a consequence of the fact that a smaller volume
of the quark-gluon phase gives rise to larger grand canonical potential
density, corresponding to a smaller pressure. For a given temperature the quark-gluon phase is thus
 less favorable (more suppressed) in a finite volume as compared to the
infinite volume limit.  The effective critical temperature $T_C^{\rm
eff}$ which can be thought of as the
temperature where the effective number of degrees of freedom increases
substantially, is strongly volume dependent (as was shown in
\cite{Spieles1998}; see \cite{Bhatta2016} for
qualitatively contrary findings within a different phenomenological model).

The suppression of small quark-gluon droplets - due to the
color-singlet constraint - also implies that quark-gluon admixtures relevant for the total system stem predominantly from 
relatively large droplets at $T \ge T_C^{\rm eff}$ . Any contribution of the quark-gluon phase to the total
equation of state must therefore be a relatively strong contribution. This is why the
softening of the phase-transition is reduced. As a result, one finite-size effect cancels the other finite-size effect.

The influence of the explicit finite-size effect of the
quark-gluon equation of state on the second order baryon number susceptibility is
certainly not negligible according to the model and should be addressed in analyses of
experimental observables. This is especially true if the
reaction volume as a hidden parameter implicitly changes in a series of
measurements, e.~g. in centrality dependence, while only $T$ or $\mu_B$ is supposed to vary. 

Both model calculations of, with and without explicit color-singlet constraint of the
quark-gluon plasma equation of state, approach the value $2/9$ expected for an ideal gas of massless quarks and gluons with two quark
flavors.

In Figure~\ref{fig:chi2} (bottom), the influence of a finite baryochemical potential is depicted. 
The qualitative behaviour with respect to system size variations appears to be similar to the $\mu_B=0$ case, i.~e. one observes a 
significant shift of the effective critical temperature $T_C^{\rm eff}$ with decreasing volume. The transition region around $T_C^{\rm eff}$ is characterized by strong variations of $\chi^B_2$ 
in a narrow temperature range, which is much more pronounced at finite $\mu_B$ than at $\mu_B=0$. For a given volume, the discontinuity 
of $\chi^B_2$ as a function of $T$ occurs at lower temperatures at finite $\mu_B$ as compared  to $\mu_B=0$. This reflects the decreasing $T_C$ with increasing $\mu_B$ 
already present in the infinite matter case.

Another quantity currently discussed extensively is the ratio of the fourth to second order susceptibility,
which is connected to the excess kurtosis 
\begin{equation}
\kappa_B = \frac{<(\delta N_B)^4>}{<(\delta N_B)^2>} - 3 \quad ,
\end{equation}
 via
\begin{equation}
\frac{\chi^B_4}{\chi^B_2} = \kappa_B\sigma_B^2 \quad.
\end{equation}

\begin{figure}[h]
\includegraphics[width=0.5\textwidth]{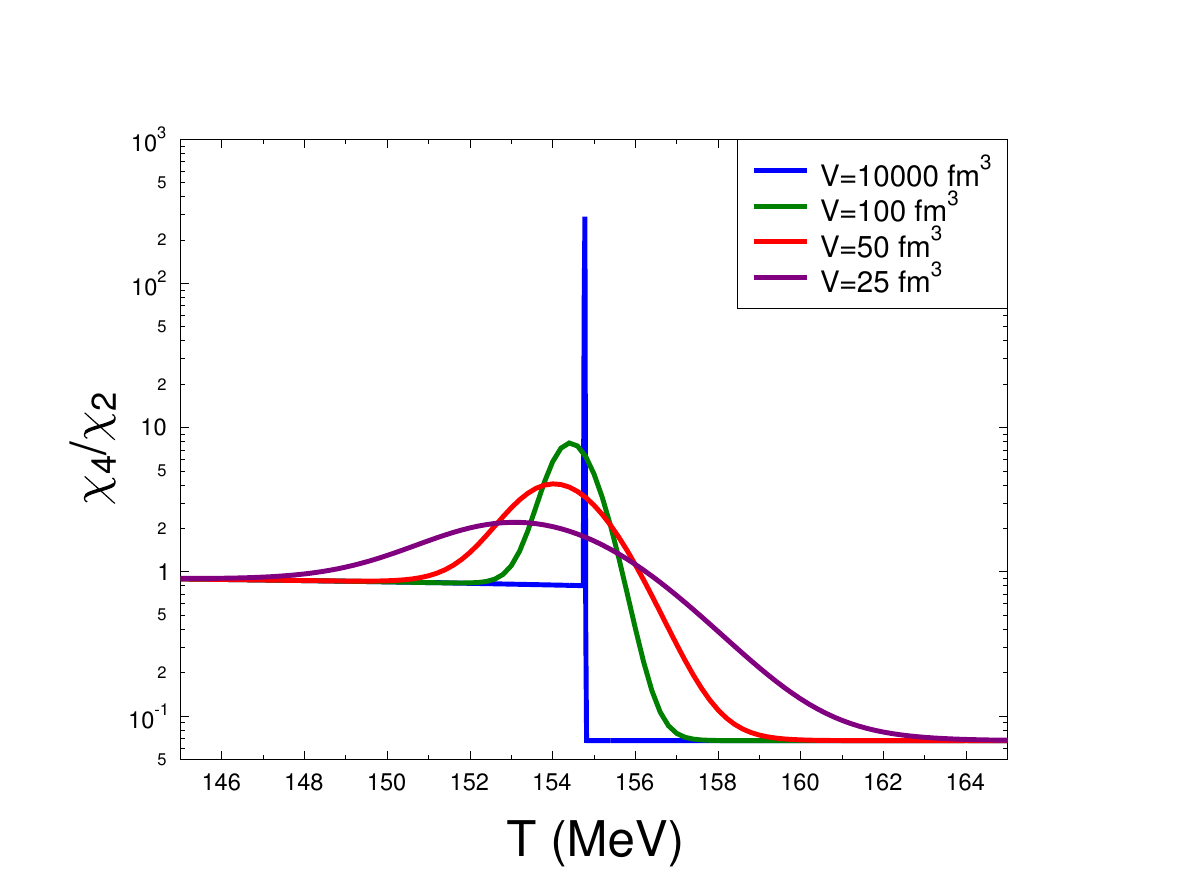}
\caption{
Fourth to second order baryon number susceptibility ratio $\chi^B_4/\chi^B_2$ as function of
temperature at $\mu_B=0$. For the QGP phase the
equation of state without explicit volume effects is employed. 
} 
\label{fig:chi4chi2inf} 
\end{figure}

This quantity is of particular interest since it is experimentally
observable, while there also exist lattice QCD results to compare with
\cite{Luo2017}\cite{Bazavov2017}. Moreover, the volume and temperature terms cancel out, when the ratio of the susceptibilities 
is used. Figure~\ref{fig:chi4chi2inf}
shows the fourth to second order baryon number susceptibility ratio $\chi^B_4/\chi^B_2$ as function of
temperature at $\mu_B=0$ for the two-phase model. Again, we consider first the
scenario without color singlet-constraint for the quark-gluon phase.
The two-phase model exhibits a strong
divergence of the susceptibility ratio with increasing volume at the critical
temperature (of infinite matter) $T_C^{\infty}$. This 
translates to extreme net-baryon number fluctuations on an
event-by-event basis in a small temperature range. However, the critical behaviour
exhibited by this model scenario comes along with the assumption of
a first order phase transition for the infinite matter limit, which is, as was stated above, not
characteristic of strongly interacting matter according to lattice QCD at $\mu_B=0$.
In any case, the divergence of $\chi^B_4/\chi^B_2$ is damped for smaller volumes
according to the model calculations. This is plausible since the absolute effect of
the system fluctuating between the two phases (or the maximum correlation
length, respectively) is limited by the finite system size.

Consistent with Figure~\ref{fig:chi2inf}, the ratio of fourth to second
order susceptibility drops from $\approx 1$
(expected for an ideal hadron gas) to $\approx 2/(3 \pi^2)$ (expected for a gas of free, massless u/d
quarks and gluons) in a relatively small temperature range around
$T_C^{\infty}$ even for small total volumes. This picture changes as soon as a
more realistic ansatz for the quark-gluon phase is chosen, i.~e., the
color-singlet constraint is preserved in the partition function.

\begin{figure}[h]
\centering
\includegraphics[width=0.45\textwidth]{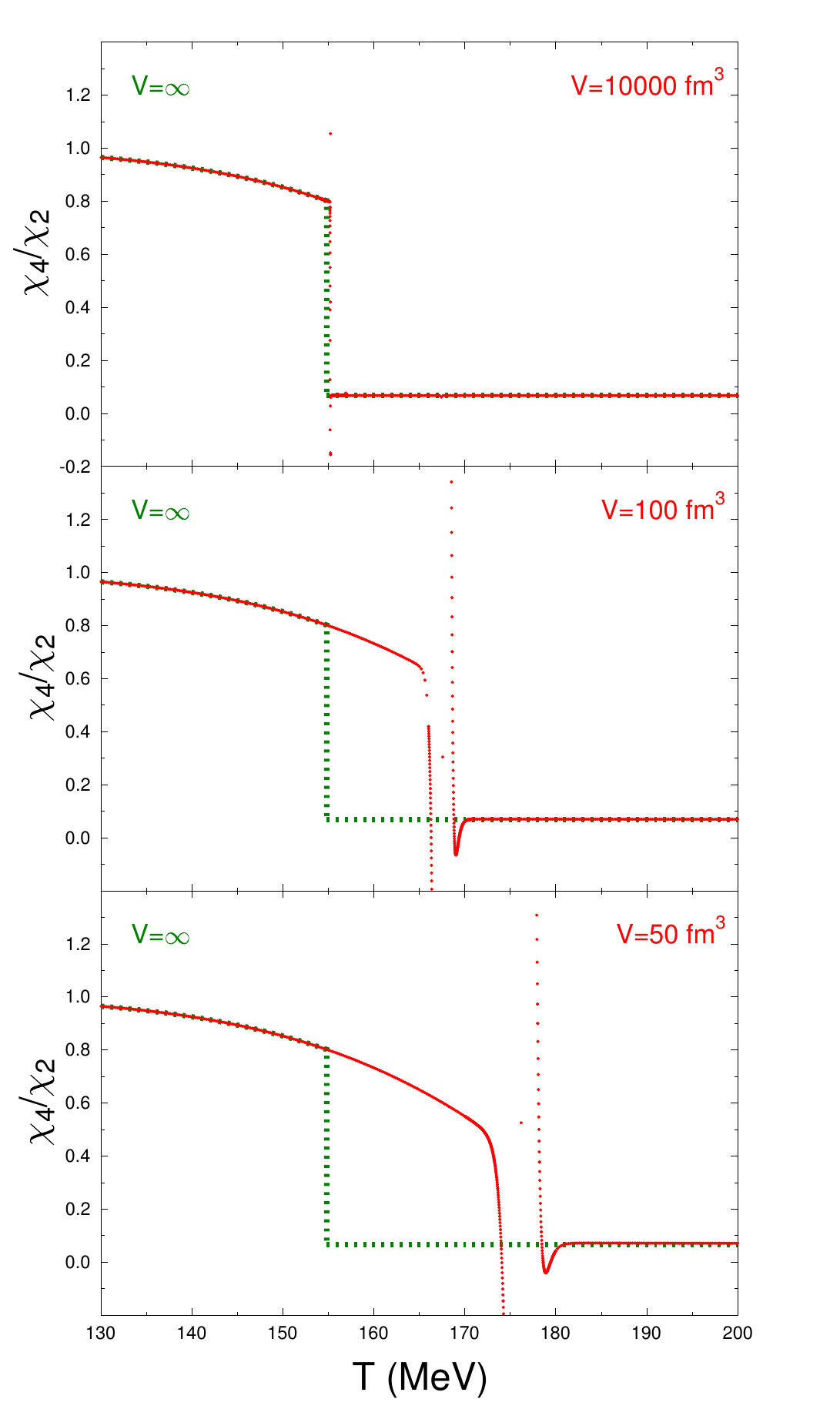}
\caption{
Fourth to second order baryon number susceptibility ratio $\chi^B_4/\chi^B_2$ as function of
temperature at $\mu_B=0$ for different volume sizes. For the QGP phase the
equation of state with explicit volume effects is employed. The green dots
show the infinite volume limit of the model.
} 
\label{fig:chi4chi2} 
\end{figure}

Figure~\ref{fig:chi4chi2}
shows the fourth to second order baryon number susceptibility ratio $\chi^B_4/\chi^B_2$ as function of
temperature at $\mu_B=0$  with the color singlet-constraint for the quark-gluon phase.
In line with Figure~\ref{fig:chi2} we observe a temperature shift of the
sudden decrease of the ratio for small system volumes. The total system
seems to behave as if it was infinite matter, however with a higher critical temperature
marking a first-order phase transition. For a system size of $V=100~{\rm
fm^3}$ there is a temperature range above $T^{\infty}_C$ between $155~{\rm MeV}<T<165~{\rm MeV}$,
where the observable susceptibilitiy ratios reflect the properties of
a "superheated" hadron
gas, not that of a quark-gluon plasma. For even smaller volumes of $V=50~{\rm fm^3}$, this region
extends to $T\approx 175~{\rm MeV}$. Moreover, according to the model, the sudden change of the
thermodynamic bulk properties in a relatively small temperature range around $T^{\rm
eff}_C$ gives rise to a near-divergent behaviour, i.~e. very large negative
and positive values of
$\chi^B_4/\chi^B_2$.
Note that for a typical volume encountered at RHIC energies ($V \approx1000~{\rm fm^3}$ 
 in one unit of rapidity) the finite-size effect is still very weak and the effective
critical temperature is shifted only by $\Delta T^{\rm eff}_C \approx 2~{\rm MeV}$.
Thus, for such a volume a direct comparison to lQCD is reliable.

\begin{figure}[h]
\centering
\includegraphics[width=0.5\textwidth]{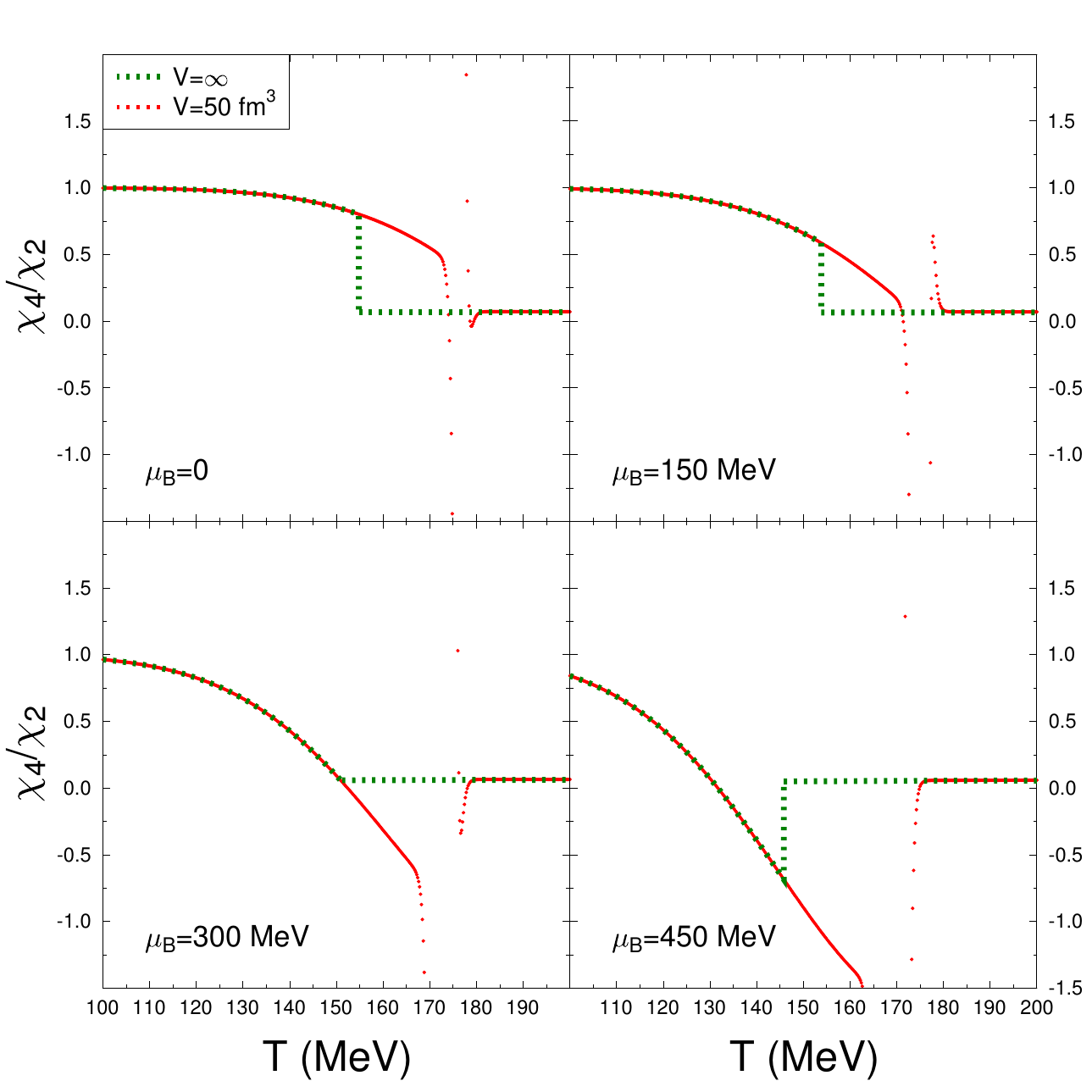}
\caption{
Fourth to second order baryon number susceptibility ratio $\chi^B_4/\chi^B_2$ as function of
temperature at different values of baryochemical potential $\mu_B$. For the QGP phase
the equation of state with explicit volume effects is employed. The results
for system size of $V=50~{\rm fm^3}$ (red dots) are compared with the infinite
volume limit (green dots).
} 
\label{fig:chi4chi2mu} 
\end{figure}

Figure~\ref{fig:chi4chi2mu}
shows the fourth to second order baryon number susceptibility ratio $\chi^B_4/\chi^B_2$ as function of
temperature for different values of  $\mu_B$ for $V=50~{\rm fm^3}$ in
comparison to infinite matter. We consider
the case with color singlet-constraint for the quark-gluon
phase.
In the infinite volume case, finite baryochemical potentials change the
susceptibility ratio as a function of temperature as compared to a system
at $\mu_B=0$. 
This is mainly due to the hadronic phase which shows strongly reduced
values of $\chi^B_4/\chi^B_2$ with increasing baryochemical potential, while the
susceptibility ratio of the pure quark-gluon phase is only weakly dependent
on $\mu_B$. The critical temperature according to the
two-phase model is lowered from $T^{\infty}_C (\mu_B=0) \approx 155~{\rm MeV}$ to
$T^{\infty}_C(\mu_B=450~{\rm MeV})\approx 147~{\rm MeV}$. At
$T^{\infty}_C(\mu_B)$, the
susceptibility ratio of the two-phase system switches from the hadron gas to the quark-gluon plasma
value. Thus, generally, it exhibits a sudden increase or decrease of its value, depending on $\mu_B$.
Only for $\mu_B=300~{\rm MeV}$, the
values of  $\chi^B_4/\chi^B_2(T^{\infty}_C)$ happen to be the same for the
hadron and the quark-gluon phase, leading to a smooth transition. At higher
values of $\mu_B$, infinite matter exhibits a negative susceptibility ratio
in a certain temperature range below  $T^{\infty}_C$, because the values of the hadronic phase are sufficiently low.

Now we consider a system of size $V=50~{\rm fm^3}$.
At a finite,
but moderate baryochemical potential of $\mu_B=150~{\rm MeV}$, one
recognizes a reduction of the susceptibility ratio in the temperature range
of the "superheated" hadron gas,
$T^{\infty}_C<T<T^{\rm eff}_C$, as compared to the $\mu_B=0$ case. Still,
the value of $\chi^B_4/\chi^B_2$ is higher in the finite system as compared
to infinite matter. In the same temperature range (where the quark-gluon contribution is
suppressed), $\chi^B_4/\chi^B_2$ drops to negative absolute values
 at $\mu_B=300~{\rm MeV}$. Here, the finite volume creates an effect contrary
to the one predicted for small baryochemical potentials, as the observable
susceptibility ratios are considerably lower than for infinite matter.  At $\mu_B=450~{\rm
MeV}$, this effect is even more pronounced. 

\section[]{Summary}

We have presented the first study on a novel finite size effect on baryon number susceptibilities. According to our schematic model, the expected change of the reaction
volume in heavy-ion collisions at different beam energies and centralities should lead to
non-trivial effects concerning $\chi^B_4/\chi^B_2$ even though one
naively expects a canceling of the (average) volume dependence in these ratios.
In order to make connections between experimental observables and the actual
thermodynamic properties of strongly interacting matter, one has to be very careful with
respect to the finite volumes of the reactions under investigation.  
In real experiments one does not
probe the phase diagram in the shape of the classic $T-\mu$ plane but in the $T-\mu-V$
space. From our model we infer that the effective equation of state of
strongly interacting matter in a finite volume is necessarily different from the
infinite matter equation of state (which can be theoretically explored with lattice QCD
calculations). 
 
We concede that the results presented here provide only a first exploratory study within a simplified model and 
can hardly be considered as quantitative predictions.
However, they point to a possibly significant complication in the analyses
of experimental observables which are aimed at comparisons with lattice QCD
predictions. It seems that more theoretical work in this respect needs to be
done.

\begin{acknowledgments}
This work was supported by the Helmholtz International Center for FAIR within the framework of the LOEWE program launched by the State of Hesse.
The computational resources were provided by the Center for Scientific Computing (CSC) of the Goethe University Frankfurt.
This work has been supported by COST Action THOR (CA15213). 
\end{acknowledgments}

\appendix
\section[]{The grand canonical potential of the two-phase system}

As described in \cite{Spieles1998}, it is assumed that the partition
function of the total system factorizes into the partition fuctions of the
two individual phases for fixed $\xi$, where $\xi$ characterizes  the
macroscopic configuration:
The volume of the hadronic phase and the quark-gluon phase are $V_h=\xi V$ and $V_q=(1-\xi) V$,
respectively. The grand canonical potential $\Phi$
of the total system in configuration $\xi$ is then given by 

\begin{eqnarray}
\Phi_\xi(T,\mu_B,V)&=& [\varphi_h(T,\mu_B,\xi V)\xi  \nonumber \\*
&+& \varphi_q(T,\mu_B,(1-\xi)V) (1-\xi)]V
\; ,
\end{eqnarray}
where $\varphi_h$ and $\varphi_q$ are the densities of the grand canonical potential of the
individual phases.\footnote{For $\mu_B=0$ - as was the case in
\cite{Spieles1998} - the grand canonical potential $\Phi$ can be replaced by the
free energy $F$ in all basic formulas of the model.}

The normalized probability for the total system being in configuration $\xi$
must then be
\begin{equation}
p(\xi;T,\mu_B,V)=\frac{\exp[-\Phi_\xi(T,\mu_B,V)/T]}{\int_0^1 \exp[-
\Phi_\xi(T,\mu_B,V)/T]d\xi} \quad .
\end{equation}

Any intensive thermodynamic quantity of the total system $A_{\rm
tot}(T,\mu_B,V)$ - including the density of the grand
canonical potential $\varphi_{\rm tot}$ - can then be expressed as

\begin{eqnarray}
A(T,\mu_B,V)&=&\int_0^1 p(\xi;T,\mu_B,V)[A_h(T,\mu_B,\xi V)\xi \nonumber \\* 
 &+& A_q (T,\mu_B,(1-\xi)V)(1-\xi)] d\xi \, .
\end{eqnarray}

\section[]{The grand canonical potential of the hadronic phase}

The model equation of state of the hadronic phase is constructed as an ideal
relativistic quantum gas of experimentally established non-strange baryon and meson resonances up to masses of
2~GeV.  We calculate the density of the grand canonical potential (which equals the negative pressure) as

\begin{equation}
\varphi_h = -\sum_i \frac{g_i}{6 \pi^2} \int_0^{\infty} \frac{dp}{E_i} \frac{p^4}{\exp [(E_i-\mu_i)/T] \pm 1}  \quad  ,
\end{equation}
where "$+$" stands for fermions and "$-$" for bosons, $g_i$ denotes the degeneracy of particle species $i$. $E_i=\sqrt{p^2+m_i^2}$ is the energy of particle species $i$ and $\mu_i$ its chemical potential. 

In order to take into account repulsive short-range interactions (or eigenvolumes) of the hadrons, all thermodynamic quantities are corrected by
the Hagedorn factor $1/(1+\epsilon/4B)$ \cite{Hagedorn1980}, where $\epsilon$ is the energy density of point particles and $B$ is the bag constant (its value is chosen consistent with Appendix~ \ref{sec:appendixqgp}). Note that $\varphi_h$ is a function of $T$ and $\mu_B$ only, it does not depend on the subsystem's volume $V$.

\section[]{The grand canonical potential of the quark-gluon plasma phase}\label{sec:appendixqgp}

The model equation of state of a color-singlet quark-gluon plasma of volume $V$,
temperature $T$ and quark-chemical potential $\mu_q = \mu_B/3$ has been derived in \cite{Elze1986}.
The deconfined phase is thought of as a gas of non-interacting quarks and
gluons in a cavity, held together by a phenomenological vacuum pressure $B$. The color neutrality and total momentum constraints on all many-particle states involved are accounted for 
with a group-theoretical projection method: Starting with a
Fock space representation of free quarks and gluons,  a group-integration over the generating
function (using an appropriate Haar measure) effectively projects out a canonical partition function
with respect to color and momentum quantum numbers. The method consistently takes
into account the discreteness of the single-particle states and thus
includes average shell effects in quark-gluon plasma droplets of
finite size. In the case of two flavors of massless quarks and fixed total momentum of zero the
resulting grand canonical partition function reads (according to  \cite{Elze1986}):
\begin{equation}\label{eq:partitionfunc}
Z(T,R,\mu_q)=\frac{1}{2}\sqrt{\frac{1}{3}\pi}\; C^{-4} D^{-3/2} \exp(-BV/T)Z_0 \quad ,
\end{equation}
where $Z_0$ is the unprojected partition function including shell corrections. It is given by
\begin{equation}
\ln Z_0(T,R,\mu_q) = X - Y \quad ,
\end{equation}
with
\begin{eqnarray}
X  &=& \pi^2 VT^3 \times  [ \frac{37}{90} +
( \frac{\mu_q}{\pi T} )^2 + \frac{1}{2} (\frac{\mu_q}{\pi T})^4 ]  \nonumber \\* 
Y &=&  \pi R T [\frac{38}{9} +2 (\frac{\mu_q}{\pi T})^2 ] 
\quad .
\end{eqnarray}
The parameters for the finite-size corrections in \ref{eq:partitionfunc} are given by
\begin{equation}
D = 2 X - \frac{1}{3}Y 
\end{equation}
and
\begin{equation}
C = 2 VT^3 [\frac{4}{3}+(\frac{\mu_q}{\pi T})^2 ] +
\frac{20}{3\pi}RT   \quad ,
\end{equation}  
where $R=(\frac{3V}{4\pi})^{1/3}$ is the radius of a spherical droplet. The
phenomenological bag pressure
has been fixed to the value $B^{1/4}=215~{\rm MeV}$ in order to recover
a critical temperature of $T^{\infty}_C\approx 155~{\rm MeV}$ at $\mu_B=0$ within the simple model (for
infinite volumes), which matches the current estimates of the chiral
transition temperature from lattice calculations (see \cite{Bazavov2017} and
references within).

The density of the grand canonical potential of the quark-gluon plasma phase is then calculated from the grand canonical partition function as
$\varphi_q (T,\mu_q, V) = -T/V \ln Z$. This quantity depends explicitely on the subsystem's volume $V$.

\end{document}